%% file: main_v7.tex
\documentclass[twocolumn,aps,pra,floatfix,superscriptaddress,notitlepage]{revtex4-1}
\usepackage[utf8]{inputenc}
\usepackage[UTF8]{ctex}
\usepackage[OT1]{fontenc}
\usepackage{amsmath}
\usepackage{amsfonts}
\usepackage{amssymb}
\usepackage{bm}
\usepackage{graphicx}
\usepackage[left=2cm,right=2cm,top=2cm,bottom=2cm]{geometry}
\usepackage{longtable,booktabs}
\usepackage{threeparttable} 
\allowdisplaybreaks
\usepackage[table]{xcolor}

\newcommand{\bnabla}{\bm{\nabla}}

\usepackage{dcolumn}
\usepackage{siunitx}
\usepackage{multirow}
\usepackage{bigdelim}

\newcommand{\balpha}{\bm{\alpha}}
\newcommand{\br}{\bm{r}}
\newcommand{\bp}{\bm{p}}

\usepackage{color}
\definecolor{BLUE}{rgb}{0.0,0.0,1.0}

\usepackage{soul}
\soulregister\cite7
\soulregister\ref7

\usepackage{float}
\ctexset{tablename=Table}

\ctexset{figurename=FIG.}

\begin{document}

\title{Nuclear-size correction to the one-loop self-energy in hydrogenlike ions}

\author{Ping~Yang~(杨萍)}
\affiliation{Innovation Academy for Precision Measurement Science and Technology, Chinese Academy of Sciences, Wuhan 430071, China
\looseness=-1}

\affiliation{University of Chinese Academy of Sciences, Beijing 100049, China
\looseness=-1}

\affiliation{Department of Physics, St.~Petersburg State University, Universitetskaya 7/9, St.~Petersburg 199034, Russia  
\looseness=-1}

\author{Aleksei~V.~Malyshev}
\affiliation{Department of Physics, St.~Petersburg State University, Universitetskaya 7/9, St.~Petersburg 199034, Russia  
\looseness=-1}

\affiliation{Petersburg Nuclear Physics Institute named by B.P. Konstantinov of National Research Center “Kurchatov Institute”, Gatchina, Leningrad region 188300, Russia
\looseness=-2}

\author{Aleksandr~N.~Kochnev}
\affiliation{Department of Physics, St.~Petersburg State University, Universitetskaya 7/9, St.~Petersburg 199034, Russia  
\looseness=-1}

\author{Vladimir~M.~Shabaev}
\affiliation{Department of Physics, St.~Petersburg State University, Universitetskaya 7/9, St.~Petersburg 199034, Russia  
\looseness=-1}

\affiliation{Petersburg Nuclear Physics Institute named by B.P. Konstantinov of National Research Center “Kurchatov Institute”, Gatchina, Leningrad region 188300, Russia
\looseness=-2}

\author{Aleksandr~S.~Shamanaev}
\affiliation{Department of Physics, St.~Petersburg State University, Universitetskaya 7/9, St.~Petersburg 199034, Russia  
\looseness=-1}

\author{Ting-Yun~Shi~(史庭云)}
\affiliation{Innovation Academy for Precision Measurement Science and Technology, Chinese Academy of Sciences, Wuhan 430071, China
\looseness=-1}

\author{Dmitry~V.~Sychkov}
\affiliation{Department of Physics, St.~Petersburg State University, Universitetskaya 7/9, St.~Petersburg 199034, Russia  
\looseness=-1}

\begin{abstract}
The nuclear-size effect on both diagonal and off-diagonal one-loop self-energy matrix elements is considered for hydrogenlike ions with $Z=60$, $82$, $90$, and $92$.
Specifically, the $1s$, $2s$, $3s$, $2p_{1/2}$, and $2p_{3/2}$ states, as well as the off-diagonal $1s-2s$, $1s-3s$, and $2s-3s$ matrix elements are considered.
The calculations are performed within the rigorous quantum-electrodynamics framework, nonperturbatively in the nuclear-strength parameter $\alpha Z$.
Excellent agreement is found with results reported in the literature.
Simple and useful approximate formulas to treat the nuclear-size correction are obtained, which, in particular, can be used to study the self-energy contributions to the field-shift factors.


\end{abstract}


\maketitle


\section{Introduction \label{sec:0}}

Quantum electrodynamics (QED) in the Furry picture~\cite{Furry:1951:115} provides a natural framework for studying highly charged ions (HCIs)~\cite{Beyer:2003:book:eng}. In this formalism, the classical electric field $V_{\rm nucl}$ induced by a nucleus is incorporated into the zeroth-order Hamiltonian, and all calculations are performed nonperturbatively in the nuclear-strength parameter $\alpha Z$ (where $\alpha$ is the fine-structure constant and $Z$ is the nuclear charge number). One can employ either the Coulomb potential for a point nucleus, $V_{\rm C}=-\alpha Z/r$, or potentials corresponding to more realistic models of extended nuclei that take into account a finite charge distribution. A comparison of the results obtained for point and extended nuclei enables one to extract the nuclear-size (NS) effect on various atomic properties.

\begin{figure}[ht]
  \centering
  \includegraphics[height=0.12\textheight]{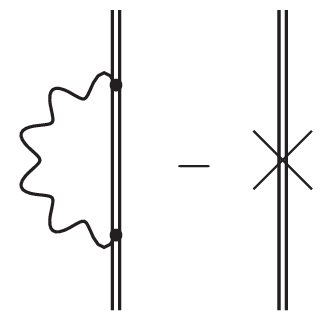}
  \caption{First-order self-energy diagram. The double line denotes the bound-electron propagator, the wavy line corresponds to the photon propagator, and the cross indicates the mass counterterm.}
  \label{fig:se_1el}
\end{figure}

Being an important correction to the electronic structure of HCIs, the NS effect becomes particularly significant when treating the corresponding isotope differences. Along with the nuclear-recoil effect, which is responsible for the so-called mass shift (MS), the difference in the nuclear-charge distribution leads to a field shift (FS). The MS and FS together determine the dominant contributions to the isotope shifts. For lighter elements, the MS gives a larger contribution, whereas for heavier elements, the FS comes to the fore. It should be noted that joint theoretical and experimental studies of isotope differences not only provide access to different nuclear parameters but also pave the way for the search for new physics \cite{Berengut:2025:119}.

The leading QED contribution to the Lamb shift of energy levels arises from the one-electron one-loop self-energy (SE) diagram shown in Fig.~\ref{fig:se_1el}. Starting from the pioneering works~\cite{Desiderio:1971:1267, Mohr:1974:26, Mohr:1974:52}, considerable progress has been achieved in all-order (in $\alpha Z$) calculations of this effect, see, e.g., Refs.~\cite{Snyderman:1991:43, Blundell:1991:R1427, Indelicato:1992:172, Cheng:1993:1817, Quiney:1993:132, Persson:1993:125, Labzowsky:1997:177, Jentschura:1999:53, Yerokhin:1999:800, Yerokhin:2005:042502, Sapirstein:2023:042804, Yerokhin:2025:012802}. The NS correction to the one-loop SE contribution has been considered in Refs.~\cite{Soff:1988:5066, Mohr:1993:158, Yerokhin:2011:012507}. In Ref. \cite{Yerokhin:2011:012507}, in particular, simple approximate formulas for this correction in the case of the $ns$ (with $n=1$, 2, and 3), $2p_{1/2}$, and $2p_{3/2}$ states in hydrogenlike uranium ($Z=92$) were derived that have proven useful, e.g., for estimating QED corrections to the FS in lithiumlike HCIs~\cite{Zubova:2014:062512}. The present work aims to extend these approximate formulas to a number of additional hydrogenlike ions ($Z=60$, 82, and 90). Moreover, we consider the off-diagonal matrix elements of the SE operator between the $s$ states. Together with the diagonal matrix elements, they can be used, e.g., to construct a modification of the model-QED operator \cite{Shabaev:2013:012513, Shabaev:2015:175:2018:69:join_pr} for treating the isotope shifts, see also Ref.~\cite{Skripnikov:2024:012807}. 

Relativistic units ($\hbar=1$ and $c=1$) and the Heaviside charge unit ($e^2=4\pi\alpha$, where $e<0$ is electron charge) are used throughout the paper.


\section{Theoretical approach and computational details \label{sec:1}}

In the present work, the nuclear-charge distribution of an extended nuclei is described by the two-parameter Fermi model:
\begin{eqnarray}
\rho(r) = \frac{\rho_0}{1+\exp{[(r-c)/a]}} \,,
\end{eqnarray}
where the parameter $a$ is fixed by the standard choice $a=2.3/(4\ln 3)$~fm, the parameter $c$ is related by the simple approximate formula
\begin{eqnarray}
c^2 = \frac{5}{3}R^2 - \frac{7}{3}a^2\pi^2 \,
\end{eqnarray}
to the root-mean-square (RMS) radius $R=\langle r^2 \rangle^{1/2}$, and $\rho_0$ is the normalization factor given by the  condition $\int \! d\bm{r} \, \rho(r)=1$. 
The potential induced by the nuclear-charge distribution $\rho(r)$ reads as
\begin{align}
\label{eq:V_F}
V_{\rm F}(r) = -4\pi\alpha Z \int\limits_0^{\infty}\! dr' \, r'^2 \frac{\rho(r')}{r_>} \,,
\end{align}
where $r_> = \max(r, r')$. 
For spherically symmetric potentials $V_{\rm C}$ or $V_{\rm F}$, one can characterize the electron state in a hydrogenlike ion by the principal quantum number $n$, relativistic angular quantum number $\kappa=(-1)^{j+l+1/2}(j+1/2)$ (where $l$ and $j$ are the orbital and total angular momenta, respectively), and the projection $\mu$ of the total angular momentum onto a chosen axis. Therefore, to zeroth order of the Furry picture, the electron obeys the Dirac equation:
\begin{align}
\label{eq:Dirac}
h_{\rm D} \psi_{n\kappa\mu} \equiv  
\left[ \balpha \cdot \bp +\beta m + V_{\rm nucl} \right] \psi_{n\kappa\mu}
=\varepsilon_{n\kappa} \psi_{n\kappa\mu}\,,
\end{align}
where $\bp=-i\bnabla$ and $\balpha$ and $\beta$ are the Dirac matrices. In the case of the point nucleus, the Dirac energies are given by the well-known Sommerfeld formula:
\begin{align}
\label{eq:Sommerfeld}
\varepsilon_{n\kappa}^{\rm C} = m \left[ 1 + \left( \frac{\alpha Z}{n_r + \gamma}\right)^2 \right]^{-1/2} \,,
\end{align}
where $n_r = n-|\kappa|$ and $\gamma=\sqrt{\kappa^2 - (\alpha Z)^2}$. For the Fermi potential~(\ref{eq:V_F}), one can solve the Dirac equation (\ref{eq:Dirac}) numerically, see, e.g., Ref. \cite{Salvat:2019:165}. Then, the NS correction to the Dirac energy is given by the difference of the corresponding Dirac eigenvalues:
\begin{align}
\label{eq:NS_dirac}
\Delta \varepsilon_{\rm NS} (n,j,l) = 
\varepsilon_{n\kappa}^{\rm F} - \varepsilon_{n\kappa}^{\rm C} \,.
\end{align} 

The renormalized one-loop SE operator reads as follows, see, e.g., Ref.~\cite{Mohr:1998:227}:
\begin{align}
\label{eq:SE_R}
\Sigma_{\rm R}(\varepsilon,\br_1,\br_2)
=&2i\alpha \int_{-\infty}^\infty \! d\omega \, D_{\mu\nu}(\omega,\br_{12})  \nonumber \\
& \times \alpha^\mu G(\varepsilon-\omega,\br_1,\br_2) \alpha^\nu - \beta \delta m \,,
\end{align}
where $D_{\mu\nu}$ is the photon propagator, $G$ is the electron Green's function (compared to Ref.~\cite{Mohr:1998:227}, we define it with the opposite sign: $G(\omega,\br_1,\br_2)=(\omega-h_{\rm D})^{-1}$), $\alpha^\mu = (1,\balpha)$, and $\delta m$ is the mass counterterm.
In order to eliminate ultraviolet (UV) divergences in Eq.~(\ref{eq:SE_R}), we employ the renormalization procedure developed in Ref.~\cite{Yerokhin:1999:800}. Namely, the UV-divergent terms are separated out and treated in momentum space, where all the divergences are covariantly regularized and explicitly canceled. The UV-finite remainder is considered in coordinate space within the partial-wave expansion for the photon and electron propagators, see, e.g., Appendix~A in Ref.~\cite{Malyshev:2024:062802}.

The diagonal and off-diagonal contributions of the SE diagram in Fig.~\ref{fig:se_1el} to an effective Hamiltonian, acting in a model space~$\Omega^{(+)}$ spanned by the one-electron positive-energy Dirac states~\cite{Shabaev:1993:4703}, can be easily derived using the two-times Green's-function method~\cite{Shabaev:2002:119, Shabaev:2024:94:inbook}. Introducing the multi-indices $a=(n_a,\kappa_a,\mu_a)$ and $b=(n_b,\kappa_b,\mu_b)$ to designate two Dirac states, one obtains the following expression for a matrix element:
\begin{align}
\label{eq:sigma_ab}
\sigma_{ab} = \frac{1}{2}& \int \! d\br_1 \! \int \! d\br_2 \, \psi^\dagger_a(\br_1)  \nonumber \\
 &\times\left[ \Sigma_{\rm R}(\varepsilon_a,\br_1,\br_2) + 
        \Sigma_{\rm R}(\varepsilon_b,\br_1,\br_2) \right] \psi_b(\br_2) \,.
\end{align}
Then, the effective SE Hamiltonian reads as
\begin{align}
\label{eq:h_se}
h_{\rm SE}(\br_1,\br_2) = \sum_{a,b}^{\varepsilon_a,\varepsilon_b>0} \psi^\dagger_a(\br_1)\, \sigma_{ab} \,\psi_b(\br_2) \,.
\end{align}
Strictly speaking, the operator~(\ref{eq:h_se}) completely describes all the one-loop SE effects within~$\Omega^{(+)}$. The diagonal matrix elements $\sigma_{aa}$ can be interpreted as the SE contributions to the Lamb shifts of the states $\psi_a$ in hydrogenlike HCIs, while the off-diagonal ones describe a non-trivial action of $h_{\rm SE}$ on one-electron functions.
We note that the operator (\ref{eq:SE_R}) conserves the angular quantum numbers, and, therefore, $\sigma_{ab} = 0$ if $(\kappa_a,\mu_a)\neq (\kappa_b,\mu_b)$. The operator~$h_{\rm SE}$, however, can hardly be used in practical calculations, e.g., in those based on the Dirac-Coulomb-Breit Hamiltonian~\cite{Faustov:1970:240, Sucher:1980:348, Mittleman:1981:1167:note}, since Eq.~(\ref{eq:h_se}) involves infinite summations over the positive-energy Dirac levels, including the continuum-spectrum states. There is a lack of sufficiently simple algorithms for the \textit{ab initio} calculations of~$\sigma_{ab}$ for such states. A naive truncation of the summations leads to an incorrect result, because the SE operator then ceases to be a short-range operator. To overcome these problems, the model-QED-operator approach was proposed in Ref.~\cite{Shabaev:2013:012513}, which preserves the short-range nature of the exact operator (\ref{eq:h_se}) and approximates it using only a finite number of the matrix elements~(\ref{eq:sigma_ab}), evaluated rigorously for low-lying Dirac bound states. For instance, the dominant contribution, corresponding to the $s$ states ($\kappa_a=\kappa_b=-1$), typically requires $\sigma_{ab}$ with $n_a \leqslant 3$ and $n_b \leqslant 3$~\cite{Shabaev:2013:012513, Shabaev:2015:175:2018:69:join_pr}. Nowadays, the model-QED operator is successfully used for the approximate SE calculations in various atomic and molecular systems~\cite{Tupitsyn:2016:253001, Pasteka:2017:023002, Machado:2018:032517, Si:2018:012504, Muller:2018:033416, Kaygorodov:2019:032505, Savelyev:2022:012806, Zaitsevskii:2023:e27077, Guo:2024:022817, Silwal:2025:042821, Lu:2025:073203}. The matrix elements~(\ref{eq:sigma_ab}) are conveniently expressed through the dimensionless function~$F_{n_a\!n_b}$ defined according to
\begin{align}
\label{eq:F_ab}
\sigma_{ab} = \frac{\alpha}{\pi} \frac{(\alpha Z)^4}{(n_an_b)^{3/2}} \,F_{n_a\!n_b}(\alpha Z) \, mc^2 \,,
\end{align}
where the factor $mc^2$ is written explicitly for clarity.

The NS correction to the SE contribution, as in the case of Eq. (\ref{eq:NS_dirac}), is determined by the difference of its values obtained for the $V_{\rm F}$ and $V_{\rm C}$ potentials,
\begin{align}
\label{eq:NS_SE}
\Delta \sigma_{\rm NS} (n_a,n_b,j,l) = 
\sigma_{ab}^{\rm F} - \sigma_{ab}^{\rm C} \,.
\end{align} 
As was noted in Refs.~\cite{Milstein:2004:022114, Yerokhin:2011:012507}, the leading dependence of $\Delta \sigma_{\rm NS}$ on $R$ and $\alpha Z$ can be factorized out in terms of $\Delta \varepsilon_{\rm NS}$. In the present work, we generalize this representation to include the off-diagonal case and express the NS correction to the SE contribution in a following symmetrized form:
\begin{align}
\label{eq:NS_G}
\Delta \sigma_{\rm NS} (n_a,n_b,j,l) =& \left[ \Delta \varepsilon_{\rm NS} \left(n_a,\frac{1}{2},l\right) 
\Delta \varepsilon_{\rm NS} \left(n_b,\frac{1}{2},l\right) \right]^{1/2} \nonumber \\
& \times \frac{\alpha}{\pi} \, G_{\rm NS}(n_a,n_b,j,l)  \,.
\end{align} 
We note that the parametrization (\ref{eq:NS_G}) involves the total NS corrections~$\Delta \varepsilon_{\rm NS}$, rather than only the leading terms of their $\alpha Z$ expansion. We also point out the use of the total angular momentum equal to $1/2$ in the prefactors on the right-hand side of Eq.~(\ref{eq:NS_G}). In particular, this is relevant for the $2p_{3/2}$ state, for which the prefactor $\Delta \varepsilon_{\rm NS}(2p_{1/2})$ is employed. With this choice of normalization, $G_{\rm NS}$ is a slowly varying function of $Z$, and its dependence on $R$ can be conveniently studied. 
Inspired by the $\alpha Z$-expansion of the function $G_{\rm NS}$, a simple three-parameter fit that includes the logarithmic term was proposed for it in Ref.~\cite{Yerokhin:2011:012507}:
\begin{equation}
\label{eq:G_approx}
G_{\rm NS}(R) = B_1 + B_2 \ln \bar{R} + B_3 \bar{R} \,,
\end{equation}
where $\bar{R} \equiv R / (1\,\text{fm})$ is dimensionless parameter equal to the value of $R$ expressed in fermi units.
The coefficients $B_1$, $B_2$, and $B_3$ depend on $Z$ and the electronic state. In Ref.~\cite{Yerokhin:2011:012507}, the $ns$ (with $n=1$, 2, and 3), $2p_{1/2}$, and $2p_{3/2}$ states for $Z=92$ were considered. In the next section, we extend this list.

Finally, since the NS effect is mainly determined by the RMS nuclear-charge radius, the FS contribution to the energy difference between two isotopes can be approximated as follows:
\begin{equation}
\Delta E_{\rm FS} = F \delta\langle r^2 \rangle \,,
\end{equation}
where $\delta \langle r^2 \rangle$ is the difference in mean-square radii and $F$ is the FS factor defined according to
\begin{equation}
\label{eq:F_factor}
F = \frac{d E(R)}{d\langle r^2\rangle} = \frac{1}{2R} \frac{d E(R)}{dR} \,.
\end{equation}
In the next section, we also compare, for $\kappa_a=\kappa_b=-1$, the values of the SE contribution to (\ref{eq:F_factor}) obtained by the numerical derivative with respect to $R$ of the results of \textit{ab initio} calculations of the SE matrix elements,
\begin{align}
\label{eq:F_1}
\Delta F_{ab}^{\rm SE}= \frac{d \,\sigma_{ab}(R)}{d\langle r^2\rangle} \,,
\end{align}
and by using the approximation given by Eqs. (\ref{eq:NS_G}) and~(\ref{eq:G_approx}),
\begin{equation}
\begin{aligned}
\label{eq:F_2}
\Delta F_{ab}^{\rm SE}\approx & \frac{\alpha}{\pi} \left[ \left(  \Delta \varepsilon_{\rm NS}^a \frac{d \varepsilon_b}{d\langle r^2\rangle} + \Delta \varepsilon_{\rm NS}^b  \frac{d \varepsilon_a}{d\langle r^2\rangle}  \right) \right. \\
&\left. \times \frac{ B_1 + B_2 \ln \bar{R} + B_3 \bar{R}}{2\sqrt{ \Delta \varepsilon_{\rm NS}^a \Delta \varepsilon_{\rm NS}^b} }  \right. \\
& \left. + \frac{\sqrt{ \Delta \varepsilon_{\rm NS}^a \Delta \varepsilon_{\rm NS}^b}}{2R}  \left( \frac{B_2}{R} + \frac{B_3}{\text{1\,fm}} \right)  \right] \,.
\end{aligned}
\end{equation}
In Eq. (\ref{eq:F_2}), we take into account that $j_a=j_b=1/2$ for the chosen value of the relativistic angular quantum number, so that the shorten notations are employed. Numerical derivatives in Eq.~(\ref{eq:F_1}) as well as for $d \varepsilon_a/d\langle r^2\rangle $ and $d \varepsilon_b/d\langle r^2\rangle $ in Eq.~(\ref{eq:F_2}) are calculated using the four-point formula,
\begin{align}
\label{eq:deriv}
\frac{d E(R)}{dR} \approx & \frac{1}{12h}  \left[ E(R-2h)-8E(R-h) \right.\nonumber \\
& \left. \qquad   +8E(R+h) -E(R+2h) \right]
\end{align}
with a typical step $h=0.1$~fm. Other step sizes and a simpler two-point formula were used to keep the accuracy of the obtained results under control.

\section{Numerical results and discussions \label{sec:2}}

In this section, we present the results of our calculations of the NS correction to the SE contributions for the $1s$, $2s$, $3s$, $2p_{1/2}$, and $2p_{3/2}$ states, as well as for the off-diagonal $1s-2s$, $1s-3s$, and $2s-3s$ SE matrix elements for hydrogenlike neodymium ($Z=60$), lead ($Z=82$), thorium ($Z=90$), and uranium ($Z=92$). The SE contributions to energy levels correspond to the diagonal case ($a=b$) in the formulas of the previous section.

\input{tables.tex}

The calculations can be briefly described as follows. We evaluate the SE contributions for both point and extended nuclei and use Eq.~(\ref{eq:NS_SE}) to obtain the NS correction. The terms in momentum space that result from the renormalization procedure are given by multi-dimensional integrals, see, e.g., Refs.~\cite{Yerokhin:1999:800, Yerokhin:2025:012802}. Their evaluation does not require the application of any partial-wave expansions and, in principle, can be performed to any desired precision. In contrast, the remainder of the SE contribution is treated in coordinate space using the partial-wave expansion. It is convenient to consider the corresponding expansion directly for the NS correction. We truncate the expansion at some $|\kappa_{\rm max}|$ for the electron propagator (typically, $|\kappa_{\rm max}|=20$--30 depending on the state and $Z$) and study the limit $|\kappa_{\rm max}|\rightarrow \infty$ by performing a polynomial (in $1/|\kappa|$) least-squares fitting of partial sums~\cite{Yerokhin:1999:800}. By trying different orders of polynomials and different data samples, an estimate of the uncertainty associated with the extrapolation procedure can be obtained. The calculations are performed in the Feynman gauge for the photon propagator.

In Table~\ref{tab:details}, an example of such calculations is given for the case of the $1s$ state in uranium. The RMS radius is chosen to be $R=5.8569$~fm, as in Ref.~\cite{Yerokhin:2011:012507}. The first line labeled ``Momentum space'' presents the results obtained for the contributions arising from the renormalization procedure. The subsequent rows provide the details of the calculations in coordinate space. Namely, the terms of the partial-wave expansion corresponding to specific values of $|\kappa|=1,2,\ldots,20$ for the electron propagator are shown. For brevity, only the sum of the terms with $|\kappa|=11$ through 20 is given. The line labeled ``$\sum_{|\kappa|>20}$[extr.]'' presents the estimate of the uncalculated part of the partial-wave-expansion series obtained by the least-squares fitting method. Finally, the total value of the NS correction for the $1s$ state is shown in the last row. Here and below, the numbers in parentheses are the uncertainties in the last digits. If no uncertainties are given, numerical values are assumed to be accurate to all digits specified.  

In Table~\ref{tab:NS_F_z92}, the NS correction to all the studied SE matrix elements for the uranium ion is shown. Our results are compared with those obtained in Ref.~\cite{Yerokhin:2011:012507}. For this purpose, we have chosen the same RMS radius as was employed there. As can be seen from Table~\ref{tab:NS_F_z92}, our theoretical predictions for the NS corrections are in excellent agreement with the previous calculations.

\begin{figure}[H]
  \centering
  \includegraphics[width=\columnwidth]{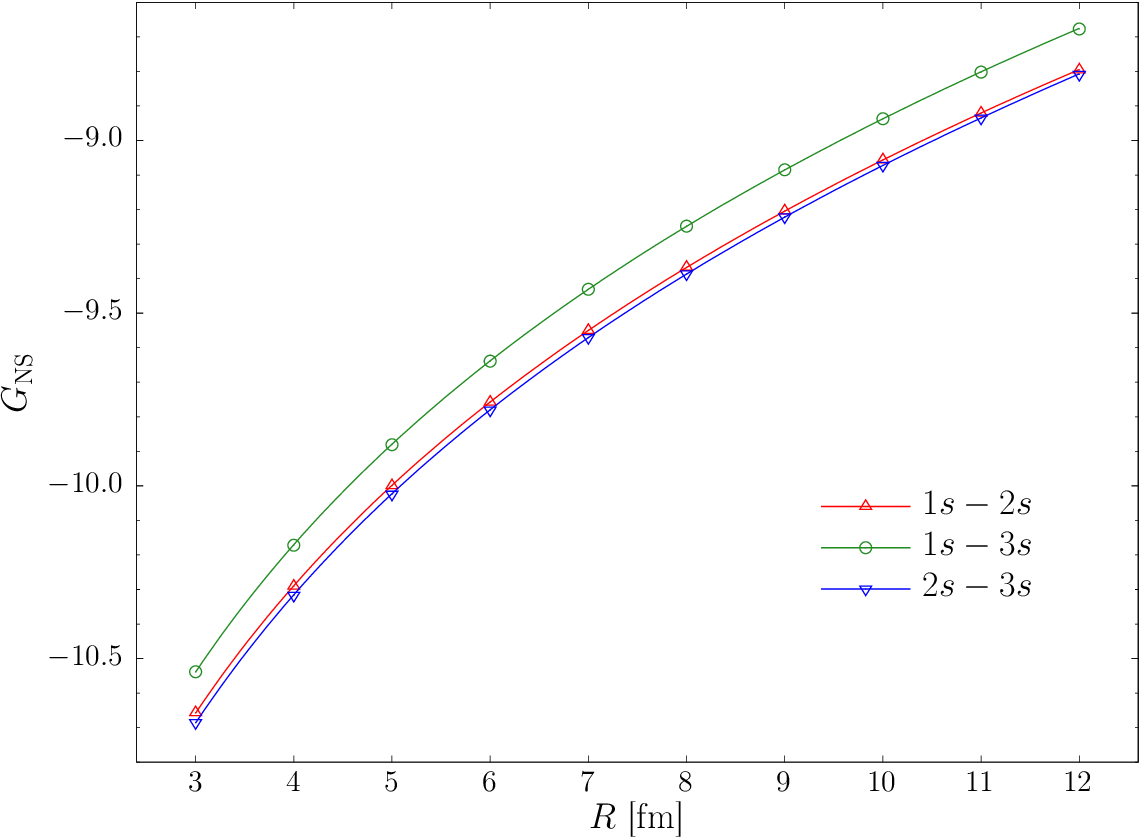}
  \caption{Nuclear-size correction $G_{\rm NS}$ defined by Eq.~(\ref{eq:NS_G}) for $Z=92$ and for the off-diagonal matrix elements of the self-energy operator as a function of the root-mean-square nuclear-charge radius $R$.}
  \label{fig:z92_off_nse}
\end{figure}

The coefficients $B_1$, $B_2$, and $B_3$ for the fit~(\ref{eq:G_approx}) of~$G_{\rm NS}$ for the diagonal and off-diagonal matrix elements of the SE operator are summarized in Tables~\ref{tab:coef_diag} and \ref{tab:coef_offdiag}, respectively. The resulting fitting functions approximate the \textit{ab initio} results in the region $R=3-12$~fm with a relative accuracy of better than $2\times 10^{-4}$. The values of the coefficients are obtained by the least-squares method applied to the results of numerical calculations in the specified range with the step $\Delta R = 1$~fm. The coefficients in Table~\ref{tab:coef_diag} for $Z=92$ are compared with those reported in Ref.~\cite{Yerokhin:2011:012507}. In general, excellent agreement is found (small deviations in the last significant digits do not affect the indicated accuracy of the approximation), with the only exception of the coefficient $B_1$ for the $3s$ state, for which the value $-11.9131$ was obtained in Ref.~\cite{Yerokhin:2011:012507}. Based on a comparison with the \textit{ab initio} results, we suppose that there was a misprint in Ref.~\cite{Yerokhin:2011:012507}. The dependence of~$G_{\rm NS}$ on the RMS radius $R$ for the off-diagonal $1s-2s$, $1s-3s$, and $2s-3s$ matrix elements in the case of $Z=92$ is shown in Fig.~\ref{fig:z92_off_nse}. One can see that the symmetrized prefactor in Eq.~(\ref{eq:NS_G}) indeed makes the functions $G_{\rm NS}$ for the different off-diagonal matrix elements close in magnitude.

Finally, in Table~\ref{tab:FS} the SE contributions to the FS factor $F$ are considered for $Z=92$ and for the diagonal and off-diagonal $s$ states. The direct numerical evaluation of the derivatives of the SE contributions with respect to $R$, Eq.~(\ref{eq:F_1}), is compared with the result obtained using the approximation for~$G_{\rm NS}$ discussed above, Eq.~(\ref{eq:F_2}). As seen from Table~\ref{tab:FS}, the approximate treatment is in reasonable agreement with the \textit{ab initio} numerical results.
Recently, experimental studies of isotope shifts have made significant progress, see, e.g., Ref.~\cite{King:2022:43}. The achieved experimental accuracy makes the corresponding measurements sensitive to the SE contributions, provided that the other contributions are evaluated with the required precision. While for HCIs rigorous QED calculations are, in principle, possible, in the case of systems with a more complex electronic structure, approximate methods must be used, e.g., those based on the model-QED-operator approach. In this context, the approximate formulas obtained in this work acquire special value.


\section{Summary \label{sec:3}}

In the present investigation, the nuclear-size correction to the self-energy contributions for the $1s$, $2s$, $3s$, $2p_{1/2}$, and $2p_{3/2}$ states as well as for the off-diagonal $1s-2s$, $1s-3s$, and $2s-3s$ matrix elements in hydrogenlike ions with $Z=60$, $82$, $90$, and $92$ is studied. Nonperturbative in the nuclear-strength parameter $\alpha Z$ calculations are performed within the rigorous QED approach, and the results are found to be in excellent agreement with those reported in the literature. Simple approximation formulas for the corresponding nuclear-size corrections are obtained. It is demonstrated that these formulas can be used to approximately treat the self-energy contribution to the field-shift factors in hydrogenlike ions.

The findings of the study show that the obtained approximation formulas, in principle, enable the extension of the model-QED-operator approach~\cite{Shabaev:2013:012513} to isotope-shift calculations in many-electron ions and even neutral atoms. This can be accomplished in several ways. First, the existing model-QED operator can be adjusted to match different values of the root-mean-squared radii~$R$. Then, one can perform the electronic-structure calculations for a set of $R$ values and, applying formulas similar to Eq.~(\ref{eq:deriv}), evaluate the field-shift factor $F$. Second, the derivative of the model-QED operator with respect to $R$ can be constructed from the very beginning, and the resulting operator, together with the original one, can also be employed to calculate $F$. The accuracy of the obtained approximate formulas is more than sufficient for treating the self-energy contribution to the isotope shifts within the model-QED approach. Both methods will be explored thoroughly in future studies. Finally, we note that, while in the present work only the $s$-state off-diagonal matrix elements are considered, the corresponding calculations can be readily extended to the $p$ states if necessary.


\section*{Acknowledgments}

This work was funded by China Scholarship Council (CSC, Grant No.~202404910570).
The work of A.~V.~M. was supported by the Foundation for the Advancement of Theoretical Physics and Mathematics BASIS (Project No.~24-1-2-74-1). 
The work of T.-Y. S. was supported by the National Natural Science Foundation of China under Grants No.~12274423, and by the Pioneer Research Project for Basic and Interdisciplinary Frontiers of Chinese Academy of Sciences under Grants No.~XDB0920101 and No.~XDB0920100.


\end{document}

%% file: tables.tex
\begin{table}[htbp]
\renewcommand{\arraystretch}{1.25}
\centering
\caption{Individual contributions to the nuclear-size correction to the self-energy contribution for the $1s$ state in hydrogenlike uranium ($Z=92$) in terms of the dimensionless function $F_{n_a\!n_b}(\alpha Z)$ ($n_a=n_b=1$ is the principal quantum number) defined in Eq.~(\ref{eq:F_ab}). The root-mean-square nuclear-charge radius is $R=5.8569$~fm. See the text for details.}
\label{tab:details}
\begin{tabular}{l@{\qquad}
                S[table-format=-1.8(1),group-separator=]
               }
\hline
\hline             
                  \rule{0pt}{1.2em}Contrib. &    
                  \multicolumn{1}{c}{Value~}                 \\            
\hline 
\rule{0pt}{1.2em}Momentum space               &    -0.01220619       \\
$|\kappa|=1$                 &    -0.00625895       \\
2                            &     0.00007930       \\
3                            &    -0.00001373       \\
4                            &    -0.00000914       \\
5                            &    -0.00000538       \\
6                            &    -0.00000333       \\
7                            &    -0.00000218       \\
8                            &    -0.00000149       \\
9                            &    -0.00000107       \\
10                           &    -0.00000079       \\
$\sum_{|\kappa|=11}^{20}$    &    -0.00000268       \\
$\sum_{|\kappa|>20}$[extr.]  &    -0.00000098(2)    \\
Total                        &    -0.01842661(2)    \\
\hline
\hline
\end{tabular}
\end{table}


\begin{table}[htbp]
\renewcommand{\arraystretch}{1.25}
\centering
\caption{Nuclear-size correction to the diagonal and off-diagonal matrix elements of the self-energy operator (\ref{eq:SE_R}) for hydrogenlike uranium ($Z=92$)
in terms of the dimensionless function $F_{n_a\!n_b}(\alpha Z)$ ($n_a$ and $n_b$ are the principal quantum numbers) defined in Eq.~(\ref{eq:F_ab}). 
The root-mean-square nuclear-charge radius is adopted to be $R=5.8569$~fm.}
\label{tab:NS_F_z92}
\begin{tabular}{c@{\qquad}
                S[table-format=-1.8(1),group-separator=]@{\qquad}
                l}
\hline
\hline
                  \rule{0pt}{1.2em}Matr. el. & 
                  \multicolumn{1}{c}{$F_{n_a\!n_b}(\alpha Z)$}                 & 
                  \multicolumn{1}{c}{Reference}   \\
\hline
  $1s$          & -0.01842661(2)    & \rule{0pt}{1.2em}This work \\
                & -0.0184266(7)     & Yerokhin~\cite{Yerokhin:2011:012507} \\
  $2s$          & -0.02898673(5)    & \rule{0pt}{1.2em}This work \\
                & -0.0289866(8)     & Yerokhin~\cite{Yerokhin:2011:012507} \\
  $3s$          & -0.0284738(4)     & \rule{0pt}{1.2em}This work \\
                & -0.0284740(6)     & Yerokhin~\cite{Yerokhin:2011:012507} \\
  $2p_{1/2}$    & -0.00247393(2)    & \rule{0pt}{1.2em}This work \\
                & -0.00247396(6)    & Yerokhin~\cite{Yerokhin:2011:012507} \\
  $2p_{3/2}$    & -0.000337183(2)   & \rule{0pt}{1.2em}This work \\
                & -0.000337183(2)   & Yerokhin~\cite{Yerokhin:2011:012507} \\
  $1s-2s$       & -0.02309386(2)    & \rule{0pt}{1.2em}This work \\
  $1s-3s$       & -0.02297341(3)    & \rule{0pt}{1.2em}This work \\
  $2s-3s$       & -0.02873565(7)    & \rule{0pt}{1.2em}This work \\
  
\hline
\hline
\end{tabular}
\end{table}


\begin{table}[htbp]
\renewcommand{\arraystretch}{1.25}
\centering
\caption{Coefficients $B_1$, $B_2$, and $B_3$ for the $ns$ (with $n=1$, 2, and 3), $2p_{1/2}$, and $2p_{3/2}$ states for the approximate evaluation of the function $G_{\rm NS}$ defined in Eq.~(\ref{eq:NS_G}) according to the simple three-parameter fit~(\ref{eq:G_approx}). All the coefficients for $Z=60$, 82, and 90 are obtained in this work.}
\label{tab:coef_diag}
\begin{tabular}{c@{\quad\,}
               l@{\quad}
               S[table-format=-2.6,group-separator=]
               S[table-format=-1.6,group-separator=]
               S[table-format=1.6,group-separator=]@{\quad}
               l}
\hline
\hline
                  \rule{0pt}{1.2em}$Z$  & 
                  State                  & 
                  \multicolumn{1}{c}{$B_1$} & 
                  \multicolumn{1}{c}{$B_2$} & 
                  \multicolumn{1}{c}{$B_3$~~~~~} &
                  \multicolumn{1}{c}{Reference}  \\
\hline
         & $1s$          & -6.2402      &  0.4301      &  0.0098  & \rule{0pt}{1.2em} \\
         & $2s$          & -6.2638      &  0.4307      &  0.0103  &   \\
$60$     & $3s$          & -6.1341      &  0.4309      &  0.0103  & This work  \\
         & $2p_{1/2}$    & -3.5873      &  0.4254      &  0.0074  &   \\
         & $2p_{3/2}$    & -0.94061     & -0.00098     &  0.00282  &   \\
\hline                                                    
         & $1s$          & -9.8022      &  0.8903      &  0.0161  & \rule{0pt}{1.2em}  \\
         & $2s$          & -10.0274     &  0.8921      &  0.0178  &   \\
$82$     & $3s$          & -9.7938      &  0.8931      &  0.0178  &  This work \\
         & $2p_{1/2}$    & -7.3085      &  0.8718      &  0.0097  &   \\
         & $2p_{3/2}$    & -1.17312     &  0.00028     &  0.00389  &   \\
\hline                                                    
         & $1s$          & -11.4304     &  1.1369      &  0.0185  & \rule{0pt}{1.2em}  \\
         & $2s$          & -11.7632     &  1.1402      &  0.0212  &   \\
$90$     & $3s$          & -11.4704     &  1.1418      &  0.0210  &  This work \\
         & $2p_{1/2}$    & -8.9565      &  1.1077      &  0.0099  &   \\
         & $2p_{3/2}$    & -1.05627     &  0.00152     &  0.00351  &   \\
\hline                                                    
\multirow{10}{*}{\raisebox{-.38ex}{$92$}}         & $1s$          & -11.8769     &  1.2084      &  0.0191  & \rule{0pt}{1.2em}This work  \\
         &               & -11.8768     &  1.2083      &  0.0191  & Yerokhin~\cite{Yerokhin:2011:012507} \\
         & $2s$          & -12.2396     &  1.2124      &  0.0220  & This work  \\
         &               & -12.2394     &  1.2124      &  0.0220  & Yerokhin~\cite{Yerokhin:2011:012507} \\
         & $3s$          & -11.9291     &  1.2143      &  0.0218  & This work  \\
         &               & -11.9131     &  1.2143      &  0.0218  & Yerokhin~\cite{Yerokhin:2011:012507} \\
         & $2p_{1/2}$    & -9.4111      &  1.1757      &  0.0098  & This work  \\
         &               & -9.4115      &  1.1759      &  0.0098  & Yerokhin~\cite{Yerokhin:2011:012507} \\
         & $2p_{3/2}$    & -1.01437     &  0.00187     &  0.00332  & This work \\
         &               & -1.0145      &  0.0019      &  0.0033   & Yerokhin~\cite{Yerokhin:2011:012507} \\
\hline
\hline
\end{tabular}
\end{table}


\begin{table}[htbp]
\renewcommand{\arraystretch}{1.25}
\centering
\caption{The same as in Table~\ref{tab:coef_diag} for the off-diagonal $1s-2s$, $1s-3s$, and $2s-3s$ matrix elements.}
\label{tab:coef_offdiag}
\begin{tabular}{c@{\quad\,}
               l@{\quad}
               S[table-format=-2.5,group-separator=]
               S[table-format=2.5,group-separator=]
               S[table-format=2.5,group-separator=]}
\hline
\hline
\rule{0pt}{1.2em}$Z$ & 
                  Matr. el. & 
                  \multicolumn{1}{c}{$B_1$} & 
                  \multicolumn{1}{c}{$B_2$} & 
                  \multicolumn{1}{c}{$B_3$}  \\
\hline
\rule{0pt}{1.2em}          & ~$1s-2s$    & -6.2676    &  0.4304   &  0.0100  \\
$60$      & ~$1s-3s$    & -6.2198    &  0.4305   &  0.0100  \\
          & ~$2s-3s$    & -6.2026    &  0.4308   &  0.0103   \\
\hline                                             
\rule{0pt}{1.2em}          & ~$1s-2s$    & -9.9188    &  0.8912   &  0.0170  \\
$82$      & ~$1s-3s$    & -9.8330    &  0.8917   &  0.0170  \\
          & ~$2s-3s$    & -9.9138    &  0.8926   &  0.0178  \\
\hline                                             
\rule{0pt}{1.2em}          & ~$1s-2s$    & -11.5913   &  1.1386   &  0.0199  \\
$90$      & ~$1s-3s$    & -11.4810   &  1.1394   &  0.0198  \\
          & ~$2s-3s$    & -11.6184   &  1.1410   &  0.0211  \\
 \hline                                            
\rule{0pt}{1.2em}          & ~$1s-2s$    & -12.0497   &  1.2105   &  0.0206  \\
$92$      & ~$1s-3s$    & -11.9316   &  1.2114   &  0.0205  \\
          & ~$2s-3s$    & -12.0852   &  1.2134   &  0.0219  \\
\hline
\hline
\end{tabular}
\end{table}


\begin{table}[htbp]
\renewcommand{\arraystretch}{1.25}
\centering
\caption{Self-energy contribution to the field-shift factor for the $ns$ (with $n=1$, 2, and 3) states and off-diagonal $1s-2s$, $1s-3s$, and $2s-3s$ matrix elements for hydrogenlike uranium ($Z=92$) in units $F_{n_a\!n_b}(\alpha Z)/$fm$^2$, where the dimensionless function $F_{n_a\!n_b}(\alpha Z)$ ($n_a$ and $n_b$ are the principal quantum numbers) is defined in Eq.~(\ref{eq:F_ab}).}
\label{tab:FS}
\begin{tabular}{c@{\qquad}
                S[table-format=-1.10(1),group-separator=]@{\qquad}
                l}
\hline
\hline
                  \rule{0pt}{1.2em}Matr. el. & 
                  \multicolumn{1}{c}{$F$}                 & 
                  \multicolumn{1}{c}{Method}   \\
\hline
  $1s$          & -0.0003557114(2)    & \rule{0pt}{1.2em}\textit{ab initio}, Eq.~(\ref{eq:F_1}) \\
                & -0.00035581         & approx., Eq.~(\ref{eq:F_2}) \\
  $2s$          & -0.0005590182(8)    & \rule{0pt}{1.2em}\textit{ab initio}, Eq.~(\ref{eq:F_1}) \\
                & -0.00055920         & approx., Eq.~(\ref{eq:F_2}) \\
  $3s$          & -0.000547100(9)     & \rule{0pt}{1.2em}\textit{ab initio}, Eq.~(\ref{eq:F_1}) \\
                & -0.00054727         & approx., Eq.~(\ref{eq:F_2}) \\
  $1s-2s$       & -0.0004455514(2)    & \rule{0pt}{1.2em}\textit{ab initio}, Eq.~(\ref{eq:F_1}) \\
                & -0.00044566         & approx., Eq.~(\ref{eq:F_2}) \\
  $1s-3s$       & -0.0004425808(5)    & \rule{0pt}{1.2em}\textit{ab initio}, Eq.~(\ref{eq:F_1}) \\
                & -0.00044269         & approx., Eq.~(\ref{eq:F_2}) \\
  $2s-3s$       & -0.000553178(2)     & \rule{0pt}{1.2em}\textit{ab initio}, Eq.~(\ref{eq:F_1}) \\
                & -0.00055333         & approx., Eq.~(\ref{eq:F_2}) \\
  
\hline
\hline
\end{tabular}
\end{table}